\documentclass[sigconf]{acmart} 
\AtBeginDocument{%
  \providecommand\BibTeX{{%
    \normalfont B\kern-0.5em{\scshape i\kern-0.25em b}\kern-0.8em\TeX}}}

\copyrightyear{2023} 
\acmYear{2023} 
\setcopyright{acmlicensed}\acmConference[CHCHI 2023]{Chinese CHI 2023}{November 13--16, 2023}{Denpasar, Bali, Indonesia}
\acmBooktitle{Chinese CHI 2023 (CHCHI 2023), November 13--16, 2023, Denpasar, Bali, Indonesia}
\acmPrice{15.00}
\acmDOI{10.1145/3629606.3629623}
\acmISBN{979-8-4007-1645-4/23/11}

\usepackage{CJKutf8}
\usepackage{multirow}
\usepackage{graphicx}
\usepackage{float} 
\usepackage{subfigure} 
\usepackage{url}
\usepackage{hyperref} 
\begin{document}

\title[Emotional Disclosure
via Diary-keeping in Quarantine]{Understanding Emotional Disclosure via Diary-keeping in Quarantine on Social Media}

\author{Yue Deng}
\affiliation{%
  \institution{Department of Computer Science and Engineering, Hong Kong University of Science and Technology}
  \city{Hong Kong SAR}
  \country{China}
}
\email{ydengbi@connect.ust.hk}

\author{Changyang He}
\affiliation{%
  \institution{Department of Computer Science and Engineering, Hong Kong University of Science and Technology}
  \city{Hong Kong SAR}
  \country{China}
}
\email{cheai@cse.ust.hk}

\author{Bo Li}
\affiliation{%
  \institution{Department of Computer Science and Engineering, Hong Kong University of Science and Technology}
  \city{Hong Kong SAR}
  \country{China}
}
\email{bli@cse.ust.hk}

\renewcommand{\shortauthors}{Deng, et al.}

\begin{abstract}
Quarantine is a widely-adopted measure during health crises caused by highly-contagious diseases like COVID-19, yet it poses critical challenges to public mental health. Given this context, emotional disclosure on social media in the form of keeping a diary emerges as a popular way for individuals to express emotions and record their mental health status. However, the exploration of emotional disclosure via diary-keeping on social media during quarantine is underexplored, understanding which could be beneficial to facilitate emotional connections and enlighten health intervention measures. Focusing on this particular form of self-disclosure, this work proposes a quantitative approach to figure out the prevalence and changing patterns of emotional disclosure during quarantine, and the possible factors contributing to the negative emotions. We collected 58, 796 posts with the "Quarantine Diary" keyword on Weibo, a popular social media website in China. Through text classification, we capture diverse emotion categories that characterize public emotion disclosure during quarantine, such as \emph{annoyed}, \emph{anxious}, \emph{boring}, \emph{happy}, \emph{hopeful} and \emph{appreciative}. Based on temporal analysis, we uncover the changing patterns of emotional disclosure from long-term perspectives and period-based perspectives (e.g., the gradual decline of all negative emotions and the upsurge of the annoyed emotion near the end of quarantine). Leveraging topic modeling, we also encapsulate the possible influencing factors of negative emotions, such as freedom restriction and solitude, and uncertainty of infection and supply. We reflect on how our findings could deepen the understanding of mental health on social media and further provide practical and design implications to mitigate mental health issues during quarantine.
\end{abstract}



%

\maketitle

\section{Introduction}

The detrimental effects on mental health caused by the COVID-19 pandemic are rampant \cite{hossain2020epidemiology, pandey2020psychological, ma2021emotional}, although it has been a long time since its first outbreak in December 2019 \cite{organization2020coronavirus}. In light of the virus transmission mechanism and the unprecedented sphere of influence, \textit{quarantine}, a direct and effective social distancing strategy, is adopted by most governments around the world \cite{wang2020subjective}. However, an emerging body of research warns of the exacerbated threat of vigorous confinement for the public's mental health under quarantine, like depression \cite{benke2020lockdown}, suicide \cite{li2022exploring}, and post-traumatic stress disorder (PTSD) \cite{hawryluck2004sars}. Understanding mental health challenges during quarantine is warranted for further research to propose supporting technologies and improve intervention measures that could help the public obtain psychological balance. 

\begin{figure}[htbp]
       \centering
    \includegraphics[width=8cm]{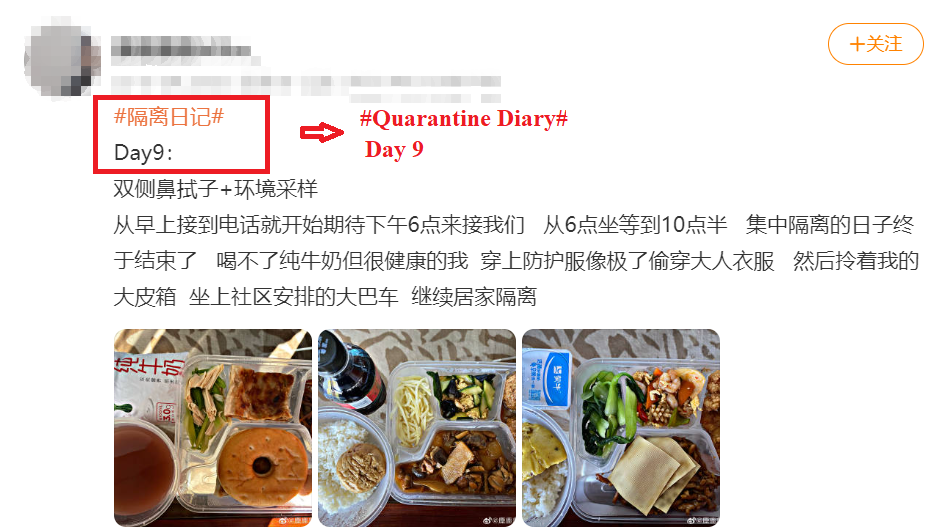}
       \caption{Example of a quarantine diary. Individuals spontaneously label the post with the hashtag \#Quarantine Diary\# and the Day No., and describe the contents of the diary (\emph{"[Nucleic acid test] Since I received the call in the morning, I have been looking forward to picking us up from 18:00 to 22:30. The days of centralized quarantine are finally over. I can’t drink milk but I am very healthy. I put on protective clothing, like stealing adult clothes. Then I carry my big suitcase, get on the bus arranged by the community, and continue the home quarantine [location]"}). }
       \label{f1}
   \end{figure}

Owing to the forced reduction in physical interaction by social distancing measures, social media has become a hub where people gather to share information and establish emotional connections \cite{alexander2014social,palen2018social}. \textit{Keeping diaries} in quarantine on social media emerges as a popular channel for citizens to share daily lives, express opinions, and vent emotions, which is a large-scale resource for understanding the mental health of individuals in quarantine. These diaries typically adhere to a specific format, recording quarantine life along with the keyword \emph{"Quarantine Diary"} and the number of days in quarantine, an example of which is shown in Figure \ref{f1}. Unlike traditional surveys, \textit{diary-keeping} on social media have distinct features: (1) proactivity. The majority of diaries are spontaneous instead of deliberately organized, featuring fragmented and piecemeal content \cite{ni2020reciprocity}. These unintentional recordings of everyday life facilitate a natural understanding of people's mental health. (2) richness. The demographic diversity on social media helps to cover different types of quarantine (e.g., home quarantine and centralized quarantine), and the measurement of mental health is not limited to psychological measurement scales \cite{canet2020longitudinal}. (3) timeliness. Timely psychological states can be reflected to help take preventive actions on mental health. Arguably, the unique characteristics of social media make it possible to comprehensively investigate the context of mental health during quarantine.

\textbf{Emotion disclosure} on social media is a prevalent expression of mental well-being \cite{gao2011harnessing,panayiotou2021coping, kelley2022using,restubog2020taking}. A rich stream of previous work has convincingly demonstrated that various emotions during the epidemic undergo changes over time \cite{feng2022gender,adikari2021emotions,guntuku2020tracking}. Unpacking the dynamics of emotional disclosure can shed light on significant implications for tracking public emotional wellness and detecting signs of distress, thereby supporting health interventions and relevant policy-making. Nevertheless, previous research has primarily focused on the overall emotional patterns throughout the entire epidemic (e.g., several months or years) \cite{ saha2020psychosocial,ashokkumar2021social} rather than quarantine, or within a specific duration of quarantine (e.g., a 14-day quarantine study) \cite{ma2021emotional}. There is still a lack of comprehensive understanding regarding how the public discloses nuanced emotions during quarantine on social media, and how these subtle emotion dimensions change over time and vary between different quarantine durations (e.g., 7-day quarantine or 14-day quarantine). Besides, the factors potentially influencing public mental well-being, particularly those contributing to distress, are also less investigated. Understanding these questions provides valuable insights for developing precise health intervention measures and informing future potential isolation situations. In this setting, the practice of continuously recording life and updating emotional status through \emph{quarantine diaries} with \emph{"Quarantine Diary"} keyword and the number of days in quarantine manifests unique values in providing temporal information of changing patterns of emotional disclosure. 





To this end, this paper explores how the public discloses emotions through diary keeping on social media during quarantine. First, identifying nuanced categories of disclosed emotions is meaningful to track public mental well-being, especially capturing potential mental health problems in quarantine. Second, given changeable emotions in quarantine, unveiling the changing patterns of emotional disclosure from the perspectives of an overall trend and multiple quarantine durations is of great importance. Third, when negative emotions have a stronger influence on behaviors and cognition of people (i.e., negativity bias theory \cite{adikari2021emotions}), deriving insight into what possible factors devoting negative emotional disclosure is warranted. As such, the following research questions are proposed:
\begin{itemize}
\item \textbf{RQ1}: What are the categories of emotional disclosure on social media in quarantine?

\item \textbf{RQ2}: What are the changing patterns of emotional disclosure through diary-keeping on social media from the perspectives of the long-term trend and various durations of quarantine?

\item \textbf{RQ3}: What factors might contribute to negative emotional disclosure on social media in quarantine?
\end{itemize}

To answer these questions, we collected 58, 796 posts from Weibo, one of the largest social media platforms, with the \emph{"Quarantine Diary"} keyword from April 29, 2020 to July 25, 2022, and performed a quantitative study. Through an open coding approach and text classifiers (RQ1), we identify six categories of emotions during quarantine (i.e., \emph{annoyed}, \emph{anxious}, \emph{boring}, \emph{appreciative}, \emph{hopeful} and \emph{happy}). By extracting temporal information with regular expression matching and statistical analysis (RQ2), we discover that posts with negative emotions occupy almost half of all posts where \emph{annoyed} was the predominant category. We further unpack the changing patterns of emotional disclosure such as the gradual decrease of negative emotions and the upsurge of \emph{annoyed} emotion near the end of quarantine. Through a topic modeling technique (RQ3), we reveal that the latent factors might account for negative emotions, such as freedom restriction and unsatisfied infrastructure for \emph{annoyed}, and uncertainty of infection and influences by management measures for \emph{anxious}. The findings shed light on health intervention measures and design implications to effectively relieve mental health problems during quarantine.



This work enriches the mental health venue in HCI community mainly by: (1) exploring prevalent emotional categories during quarantine; (2) uncovering the changing patterns of emotional disclosure from both long-term and period-based views; (3) discovering the possible factors might contributing to negative emotions; (4) proposing design implications to facilitate emotional tracking during quarantine; (5) substantiating the feasibility and importance of performing emotion analysis in the field of affective computing by a bottom-up approach that focuses on identifying context-specific emotions, rather than relying on generic sentiment analysis tools. Our study reveals emotional features and provides insights into the future direction for specific health intervention measures.
    
\section{Related Work}
\subsection{The Role of Social Media in Unveiling Mental Health During Quarantine}
Quarantine, as a restriction on the movement of people, is intended to prevent the spread of disease. To contain COVID-19, unprecedented quarantine strategies are operated and dynamically adjusted corresponding to the pandemic development. With the epidemic panic, the confinement policy tends to exacerbate mental health problems \cite{bavel2020using,ma2021emotional}. In this vein, social media platforms have become popular sites for individuals to express their opinions, describe their lives and interact with social networks, thus mapping individuals' responses and behaviors to their psychological traits \cite{hung2020social,ang2015predicting}. Owing to the prevalence of social media, researchers have paid great attention to how social media help disclose mental health, like discovering the shifts of suicidal ideation \cite{de2016discovering}, understanding the virtual attributes of mental health \cite{manikonda2017modeling} and proposing Mental Well-being Index drawing on social media \cite{bagroy2017social}. Compared with traditional methods in mental health (e.g., questionnaires and scales), social media could naturally, expediently, and timely collect emotional and behavioral information, becoming a promising place for recognizing psychological effects \cite{adikari2021emotions, lu2020psychological}. Although social media expressions cannot guarantee the identical emotional intensity or absolute authenticity of feelings with privacy or sensitivity concerns \cite{bazarova2014self}, they showcase users' thoughts, emotions, and experiences to some extent. 

Hence, this work explores mental well-being during quarantine by comprehensively understanding the characteristics of emotional disclosure via social media (RQ1), changing patterns of emotional disclosure (RQ2) and specific concerns of the public which provide guidelines to mitigate mental health disorders (RQ3). 

\subsection{Emotional Disclosure on Social Media in Quarantine}

Social media has become a widely used channel to express emotions \cite{he2021beyond,kelley2022using,de2013predicting,de2016discovering}. A burgeoning body of research investigated how emotions were disclosed with the extensive use of social media during quarantine. One strand of work focused on emotional differences between various groups of individuals \cite{wang2020subjective, zhu2021effect}. For example, Zhu et al. revealed the anxiety difference between six types of quarantine (e.g., forced quarantine at home and centralized quarantine) \cite{zhu2021effect}. Another stream of research centered on the changes of emotions where interesting changing patterns usually hide \cite{cao2021analysis, lu2020psychological,feng2022gender}. For example, Cao et al. uncovered emotional evolution during each phase of Wuhan lockdown including the incubation phase, explosive phase, declining phase and stable phase \cite{cao2021analysis}.


In previous research the measurement of emotions was mostly based on existing API \cite{li2022exploring,hung2020social}, psychological scale \cite{canet2020longitudinal}, or only from the perspective of words and themes \cite{arora2021role} rather than scenario-based analysis. However, emotions are comprehensive manifestations of not only words but also contexts. Furthermore, the majority of previous studies about emotion categories were theory-driven, e.g., drawing on Plutchik’s model with basic emotions \cite{adikari2021emotions} or LIWC with various emotional dimensions \cite{ashokkumar2021social} to determine emotion categories. 

The specific emotions expressed by individuals in quarantine, utilizing a bottom-up approach to emotion categorization, remain yet to be established. Thereby, this study contributes to this area by employing an inductive approach to investigate common emotions that align closely with the context of quarantine.

\subsection{Online Diaries on Social Media in Quarantine}
An online diary, also known as a web diary, is defined as a personal journal or diary that individuals choose to publish on the World Wide Web \footnote{https://en.wikipedia.org/wiki/Online\_diary}. The United States witnessed the emergence of the earliest online diaries around 1994 \cite{ben2020diary}. In China, the rise of online diary-writing began in 2000 when Lu Youqing's cancer battle diary was serialized on a popular online literature website, attracting a significant number of followers \cite{hockx2015internet}. The introduction of blogs to China in 2002 further popularized online diary-writing, making it widely recognized. Thus, online diaries have a significant and extensive history, serving as a common form for people to document their daily lives. During the lockdown in Wuhan, residents turned to social media like WeChat and Weibo, and started writing and sharing their experiences via "Lockdown Diaries", providing a glimpse into life in a sealed city \cite{ni2020reciprocity}. The diaries attracted scholarly attention, with studies focusing on various aspects such as feminist issues \cite{bao2020three}, and the impact of the pandemic on daily life and work. They were part of a global trend, with pandemic diaries appearing worldwide, offering access to wider audiences and facilitating interaction \cite{nytimes2020quarantine}. However, this online exposure also brought unwanted attention and censorship. Previous studies primarily focused on Wuhan, including lockdown diaries in Wuhan or nationwide quarantine diaries during the outbreak period in Wuhan \cite{feng2022gender}. Nevertheless, this diary form exists nationwide, and the quarantine measures have been in place for an extended period of time.

In general, diaries that convey emotional disclosure across the country during the extended period of quarantine remain unexplored. This research gathers nationwide data spanning approximately 2 years to uncover the changing patterns of emotional disclosure as quarantine becomes a new normal, as well as to identify potential factors associated with negative emotions.

\section{Method}
\begin{figure*}[htbp]
\centering 
\includegraphics[width=0.82\textwidth]{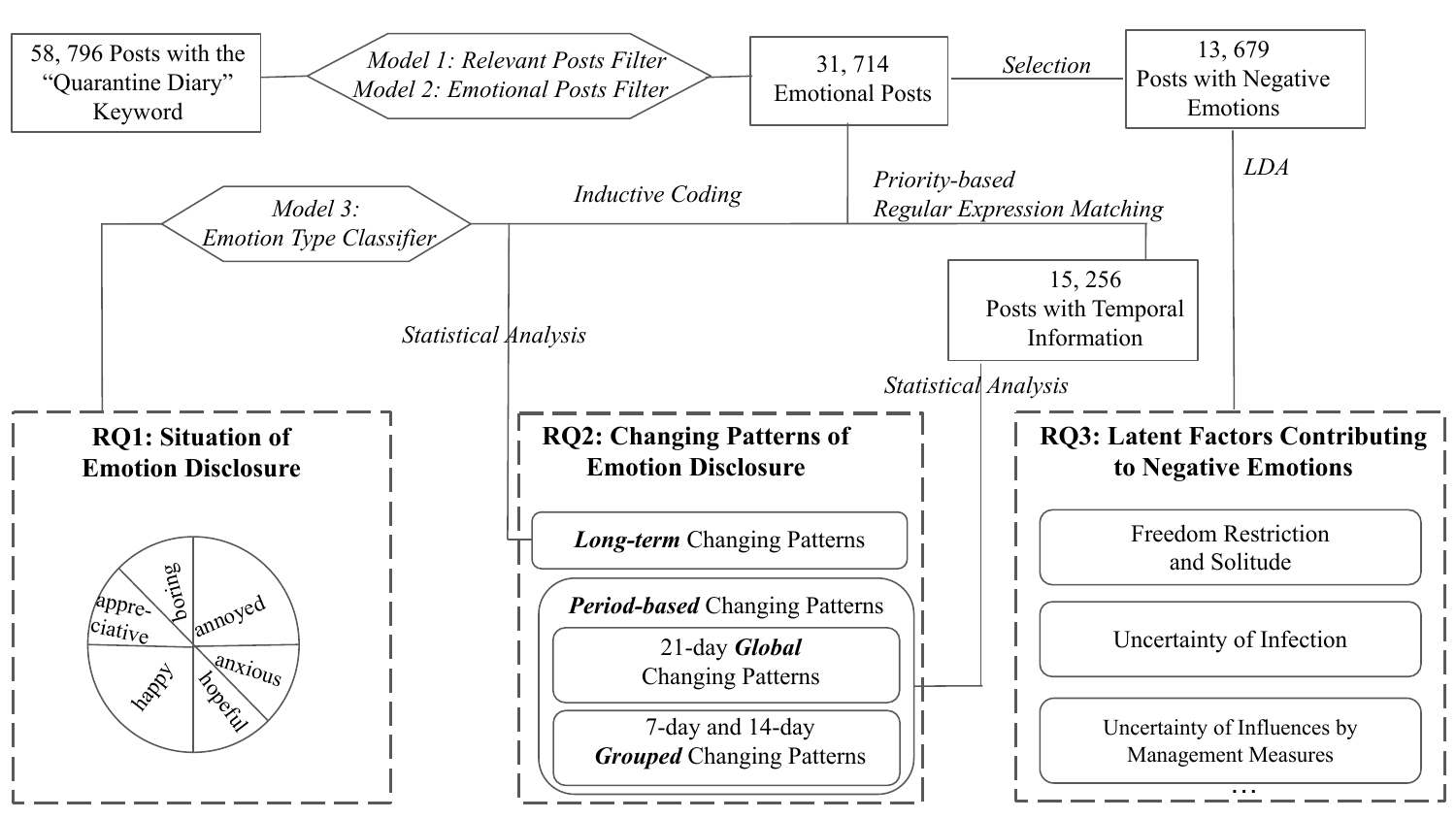} 
\caption{The overall flow chart to understand the emotional disclosure during quarantine.} 
\label{flowchart}
\end{figure*}

We adopted a quantitative study to comprehensively understand emotional disclosure during quarantine. More specifically, we first described the process of data selection and collection (Section \ref{Data Selection and Collection}) to contextualize our scenario. Then we applied inductive coding and text classifiers to comprehend the emotions during quarantine (RQ1, Section \ref{RQ1}), utilized regex-matching-based statistical analysis to unveil the changing process of emotions (RQ2, Section \ref{RQ2}), and leveraged topic modeling to unpack the possible related factors affecting negative emotions (RQ3, Section \ref{RQ3}). Figure \ref{flowchart} exhibits the overall flowchart.

\subsection{Data Selection and Collection}\label{Data Selection and Collection}
Weibo, one of the largest microblogging websites in China similar to Twitter, has become a dominant site for understanding the well-being and health trends of the public due to its characteristics of sharing, disseminating and receiving instant information \cite{li2020impact, chen2021exploring, he2022help}. Weibo was a significant and popular channel for recording daily life and expressing feelings during quarantine, where posts with hashtag "\emph{\#Quarantine Diary\#}" had owned more than 1.2 billion reads and more than 0.5 million discussions (including original posts and retweets) \footnote{\url{https://m.s.weibo.com/vtopic/detail\_new?click\_from=searchpc\&q=\%23\%E9\%9A\%94\%E7\%A6\%BB\%E6\%97\%A5\%E8\%AE\%B0\%23}}.

To obtain a preliminary understanding of the emotional status in quarantine, we kept track of posts related to quarantine on Weibo in advance. We found users applied some strategies to identify topic-related posts including hashtags, Super Topic (i.e., communities consisting of people having common interests and following specific topics \cite{supertopic}) and keywords. In order to make the data comprehensive, we selected the keyword-based collection approach rather than other approaches based on hashtags or Super Topic. Moreover, the most straightforward keyword was "\emph{Quarantine}", but posts with "\emph{Quarantine}" were miscellaneous containing official quarantine policies, suggestions after quarantine or even advertisements of primers (i.e., cosmetics). Compared with "\emph{Quarantine}", we noticed that posts with the "\emph{Quarantine Diary}" keyword, where users tended to record and share their activities and emotions for one day or several days, getting closer to the personal quarantine lives and self-expression. Hence, we chose "\emph{Quarantine Diary}" as the keyword, so as to make the collected data align with our scenario.
 
We collected the data with the assistance of the WeiboSuperSpider tool \cite{WeiboSuperSpider} ranging from April 29, 2020, to July 25, 2022. Since we only considered the quarantine in China, we removed the overseas posts (i.e., the quarantine locations are not in China) by identifying the keyword "\emph{Overseas Quarantine}" and the locations of posts in self-identified profiles.

In total, 58, 796 original posts were retrieved with the keyword "\emph{Quarantine Diary}" from 34, 677 distinct users. 

\subsection{RQ1: Understanding Emotional Disclosure Categories}\label{RQ1}
This section described how we constructed two filtering models for obtaining emotional posts and one classification model for emotional categories, to understand the categories of emotional disclosure in quarantine. 
\subsubsection{Filtering}
\label{section3.2.1}
Although we had chosen a suitable keyword, it was inevitable that the data might contain a substantial amount of noise. After manually analyzing a random set of 200 posts (saturation was achieved after coding 80-100 posts), we found the majority of such noisy posts could be divided into two types: (1) irrelevant to the scenario (e.g., posts published by quarantine volunteers and fans caring about their superstars in quarantine, and official blogs or news); (2) relevant but emotionless such as posts with mild and neutral attitudes (e.g., \emph{"take things as they come"}), and those without obvious emotions (e.g., \emph{"In the morning, I had a COVID-19 test, took the blood pressure and did my homework."}).

To remove these noisy data, we constructed two deep-learning-based models. Model 1 (relevant posts filter) was a text classifier determining whether posts were related to the quarantine context and Model 2 (emotional posts filter) decided whether posts expressed emotions. To prepare the training data, two authors first manually coded a random sample of 200 posts independently. Cohen's Kappa, a statistical measure for evaluating agreement between raters in qualitative items, is favored over percent agreement due to its consideration of chance agreement likelihood \cite{mchugh2012interrater}. A significant level of agreement was achieved between the two raters (Model 1: Cohen's Kappa: 0.89; Model 2: Cohen's Kappa: 0.90). Following their discussion, the two coders successfully reached a consensus on annotations. Then, each annotator separately coded another 400 posts, reaching a total of 1, 000 training samples for Model 1 and 781 training samples for Model 2, since some posts might be irrelevant among the 1, 000 samples.
Based on the training datasets, we tried traditional machine-learning-based methods (SVM \cite{cristianini2000introduction}and XGBoost \cite{chen2016xgboost}) and deep-learning-based methods (GRU \cite{cho2014learning}, LSTM \cite{hochreiter1997long}, and BERT \cite{devlin2018bert}) to build our filters. In order to prevent overfitting due to the limited labeled data, we utilized Adam as an optimizer and set the parameter of the Dropout layer as 0.2 during the training of deep learning models. BERT was fine-tuned based on BERT-wwm whose corpus was Chinese Wikipedia \cite{cui2021pre}. Finally, BERT achieved the best performance for both Model 1 (F1 score: 95.4\%) and Model 2 (F1 score: 88.1\%). The two models were applied to filter all posts, so the remaining posts were probably emotional.

\subsubsection{Emotion Category Identification and Classification} 

In order to investigate common emotions that align closely with the context of quarantine and considering the limited effectiveness of existing sentiment analysis models, we manually constructed a codebook of emotional categories by corpus annotation, and developed a multi-class text classifier to quantify posts with various emotions specifically related to the "\emph{Quarantine Diary}" keyword.

We first employed an open coding approach \cite{flick2022introduction} to identify the common emotions in quarantine. By taking this inductive approach, we allowed the codes to emerge naturally from the analysis. Specifically, two authors conducted manual coding of 200 randomly selected posts independently (saturation was achieved after coding about 100 posts) and generated initial codes from a nuanced perspective (e.g., "\emph{joy}", "\emph{extremely happy}", "\emph{worried}"). Subsequently, several rounds of meetings, comparisons, and discussions were held to merge the initial codes, and finally generated a codebook with high-level emotional categories (e.g., "\emph{happy}" and "\emph{anxious}"). Through the data-driven inductive coding, there were six categories of emotions where \emph{annoyed}, \emph{anxious} and \emph{boring} reflected the negative emotional disclosure, and \emph{appreciative}, \emph{hopeful} and \emph{happy} captured the positive emotion characteristics, shown in Table \ref{table1}. 

To provide a quantitative description of emotional posts and prepare for subsequent analysis, it was necessary to determine which emotion was specifically expressed in each post. The open coding revealed that the majority of posts contained a single emotion, with only a very small number exhibiting mixed emotions, often with one primary emotion. With this understanding, we developed Model 3, a single-label emotion classifier to identify the emotion categories within these potentially emotional posts. The annotation process of the training dataset is similar to Model 1 and Model 2, and the annotation data is the 1, 000 unfiltered training samples in Model 1 (Model 3: Cohen's Kappa: 0.79). Coding unfiltered samples could simultaneously validate the manual annotation accuracy for two previous models. To enhance the classification performance, one of the authors manually coded an additional 1, 500 samples, resulting in a total of 1, 237 training samples. Due to the outstanding performance of BERT in Section \ref{section3.2.1}, we also selected BERT as our classifier. We fine-tuned BERT with RoBERTa-wwm-ext-large \cite{cui2021pre} based on the complexity of multi-class classification (F1 score: 66.13\%, with the similar level of performance to other works in multiclass classification of emotions on social media \cite{bhende2022integrating, 4427100, elfaik2021social,rahman2023multi}). Then, we applied the classification model to 31, 714 potentially emotional posts.


\subsection{RQ2: Investigating the Changing Patterns of Emotional Disclosure}\label{RQ2} 
In this section, we described how we identified the temporal information in posts and investigated temporal characteristics from long-term and period-based views.

Data utilized for this section was 31, 714 emotional posts. After browsing these emotional posts, we found the temporal information of posts would be expressed in three ways as follows: (1) "\emph{Day}" or "\emph{D}" plus a number (e.g., "\emph{Day 9}" or "\emph{D 9}"); (2) using an ordinal number (e.g., "\emph{the first day}"); (3) "Quarantine Diary" plus a number (e.g., "\emph{Quarantine Diary 9}"). Moreover, there was multiple temporal information in some posts containing two or three types of temporal information or utilizing one type of temporal information more than once. To circumvent this challenge, we adopted a priority-based regular expression matching approach to extract the temporal information for each post. The priority was the same as the order mentioned above. If there was more than one temporal number in one type, we selected the maximum number as the temporal information. Note that the priority order and the maximum temporal approach were determined based on the accuracy of temporal information after manual coding. After the extraction operation, there existed 15, 256 posts with temporal information. 

To comprehensively explore how emotional status changed during quarantine, we considered it from two views: (1) long-term changing patterns, the temporal characteristics of emotions from April 2020 to July 2022 to have an overall exploration of emotional disclosure; (2) period-based changing patterns, the temporal characteristics of emotions related to quarantine durations to further understand the day-by-day emotions, including 21-day global changing patterns (i.e., based on posts whose temporal information was within 21 days, e.g., "\emph{Quarantine Diary 6}"), as well as 7-day and 14-day grouped changing patterns (i.e., based on users whose quarantine duration were 7 days or 14 days). Note that 7 days and 14 days were two typical quarantine periods adopted by the government in China \cite{policy} and recommended by WHO \cite{world2020considerations}. We used proportions instead of numbers to capture changing patterns of emotional disclosure. By doing so, it allows us to effectively unveil relative distribution changes among emotional categories, considering the varying number of posts collected at different points in time.

\subsection{RQ3: Exploring Possible Factors Contributing to Negative Emotions}\label{RQ3}
The negativity bias theory \cite{adikari2021emotions} indicates that negative emotions play a more significant role in the public's behaviors and cognition. In the context of quarantine, \emph{annoyed} and \emph{anxious} generally exhibit a stronger expression of personal concerns and potentially have more direct impacts on emotional well-being, relationships, and life quality, while \emph{boring} typically represents a mundane account of daily life. Hence, in this section, we described how we explored factors that might lead to \emph{annoyed} and \emph{anxious}.

\subsubsection{Topic Model Building}
To keep track of what individuals disclose about their quarantine lives, we conducted a topic modeling technique to characterize the topics of post contents. Specifically, we adopted Latent Dirichlet Allocation (LDA) \cite{blei2003latent}, a type of classic and useful unsupervised topic extraction model, to explore potential and abstract topics in a large number of posts where each topic was composed of a set of frequent keywords. LDA was a widely-adopted approach for understanding social media data by prior work \cite{li2021hello,jhaver2019does,pennacchiotti2011investigating}. The first exploratory attempt of LDA suggested that there still remained a lot of meaningless or irrelevant keywords in each topic \begin{CJK}{UTF8}{gbsn}(e.g., 像是-seem, 感觉-feel, 我会-I will).\end{CJK} To make keywords more amenable and reliable for analysis, we applied a two-iteration LDA framework that updated stopwords during the process. LDA was applied to 11, 525 \emph{annoyed} posts and 2, 154 \emph{anxious} posts, respectively. The details were as below:
\newpage
\textbf{Discovering context stopwords (iteration 1):}
\begin{itemize}
\item \textbf{Preprocessing:} First we implemented a noise removal operation to delete URLs and punctuations, and convert traditional characters. Jieba, a widely used tokenization tool, was applied to segment Chinese texts \cite{jieba}. We chose the HIT Chinese stopwords table as the basic stopwords set \cite{hit}. Then we also further filtered words by taking into account part of speech and frequency of words. In particular, we only retained nouns and verbs after tokenization which could show clearer and more explicit meanings and help authors to infer the topics. Moreover, we removed extreme low-frequency words (i.e., words whose frequency was smaller than 5) and extreme high-frequency words (i.e., occurrence proportion was more than 0.5). Words with extremely high frequency suggested that they might lack representativeness. We also built bigrams to represent two frequently co-occurring words \begin{CJK}{UTF8}{gbsn}(e.g., 核酸\_检测-nucleic acid test)\end{CJK}.

\item \textbf{Hyperparameter tuning:} The most significant parameter in LDA was the number of topics k. We chose k based on the coherence score $C_{v}$, a widely-adopted coherence metric \cite{roder2015exploring}. Consequently, we set the parameter space between 2 and 15. With a systematic grid search of the range, the optimal value of k was 6 and 12 for \emph{annoyed} and \emph{anxious}, respectively. Then, we applied k-topic LDA and output 50 keywords for each topic.

\item \textbf{Stopwords updating:} In our specific context, we included words that lacked clear and evident implications for human interpretation into the basic stopwords set. Additionally, we combined the high-frequency keywords, which appeared in more than one-third of all topics, with the basic set.


\end{itemize}

\textbf{Unveiling topic clusters (iteration 2):} We added the context stopwords into the basic stopwords set and other steps of preprocessing and hyperparameter tuning remained unchanged. The optimal numbers of topics k were 15 and 14 for \emph{annoyed} and \emph{anxious}, respectively, and then we ran LDA again. 

\subsubsection{Interpretation of Topics with Thematic Analyasis}
The results of LDA were sets of representative keywords for topics. We tried to figure out discussion themes for \emph{annoyed} and \emph{anxious} posts. Therefore, we applied an inductive thematic analysis \cite{braun2012thematic} to annotate each topic based on the understanding of posts and human interpretation. Specifically, two authors, who had experienced quarantine themselves and comprehended the scenario, individually distributed theme labels to topics and generated initial labels based on these sets of keywords and original posts. Through several rounds of comparison and discussion, they merged some topics describing similar themes, deleted those containing miscellaneous information, and inductively concluded final topical themes.

\section{FINDINGS}
In this section, we reported the findings on the categories of emotional disclosure in quarantine. In Section \ref{section4.1}, we unveiled the willingness of people to express various emotions on social media during quarantine. In Section \ref{section4.2}, we described how emotional status changed from the long-term and period-based perspectives to probe into the changing patterns. In Section \ref{section4.3}, we discovered the latent factors that might give rise to negative emotions. The findings elucidated mental health challenges of quarantine (RQ1), the evolutionary process of emotional disclosure (RQ2) and potential influencing factors corresponding to negative affections (RQ3).
 
\subsection{RQ1: Emotional Disclosure Categories}
\label{section4.1}

\begin{figure*}[htbp]
\centering 
\includegraphics[width=0.9\textwidth]{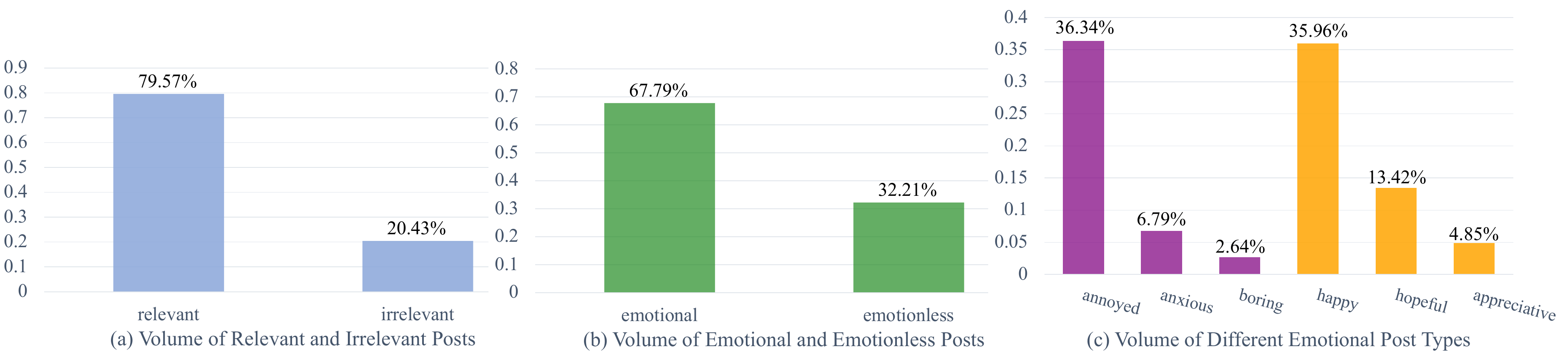} 
\caption{The distribution of (a) relevant posts, (b) emotional posts and (c) six categories of emotional posts. The results indicated negative emotions were non-negligible.}
\label{figure2}
\end{figure*}

The statistical results of emotional disclosure during quarantine were demonstrated in Figure \ref{figure2}. Based on the results of three models constructed in Section \ref{RQ1}, we concluded the following findings: (1) Most posts (79.57\%, N=46, 785) containing the "\emph{Quarantine Diary}" keyword tallied with our research context shown in Figure \ref{figure2} (a), proving that this type of data collection approach was feasible; (2) Emotional posts were predominant (67.78\%, N=31, 714) compared with emotionless ones indicated in Figure \ref{figure2} (b), which demonstrated that people in quarantine were willing to express their emotions on social media; In Figure \ref{figure2} (c), (3) Posts with negative emotions (45.77\%, N=14, 516) in quarantine occupied almost the half of all emotional posts; (4) \emph{Annoyed} (36.34\%, N=11, 525) was the dominant category in amount compared with \emph{anxious} (6.79\%, N=2, 154) and \emph{boring} (2.64\%, N=837); (5) In terms of positive emotions, \emph{happy} (35.96\%, N=11, 405), \emph{hopeful} (13.42\%, N=4, 256) and \emph{appreciative} (4.85\%, N=1, 537) were sorted in descending order. 
 \vspace{3pt}
\begin{table*}[htbp]
\caption{The codebook of emotional categories with the "Quarantine Diary" keyword.}
\label{table1}
\centering
\begin{tabular}{p{2cm}|p{2cm}p{5cm}p{4cm}}
   \toprule
   Polarity & Category & Description & Example\\
   \midrule
   \multirow{3}{1.5cm}[-3.5em]{Negative} 
   & \textbf{annoyed} & a feeling of irritation and anger. It also includes complaining and feeling imprisoned.  & \emph{There is a time limit even in jail. When can we get out?} \\
        \cline{3-4}
		& \textbf{anxious} & a feeling of uneasiness and worry. Being nervous about health, food and
   situation. & \emph{My roommate has a fever and sometimes coughs. I don’t know when I can get a second test. I’m really scared.} \\
        \cline{3-4}
        & \textbf{boring} & a feeling of no interest and tediousness. Being numb, tired or negatively describing trivia. & \emph{I feel like I'm in a cycle every day.} \\
    \midrule
    \multirow{3}{1.5cm}[-3.5em]{Positive} 
    & \textbf{appreciative} & a feeling of gratitude or pleasure. It includes gratitude to servers, the government and families. & \emph{I received vegetables and eggs from the government. Thanks to our country and the staff.} \\
        \cline{3-4}
		~ & \textbf{hopeful} & a feeling of optimism about the future. It contains being hopeful about themselves and the epidemic. & \emph{Only one day left! come on!} \\
        \cline{3-4}
        ~ & \textbf{happy} & a feeling of pleasure or contentment. It also includes being relaxed and positively describing daily life. & \emph{I am happy today. I drank a cup of milk tea, and the egg-fried rice tasted good.} \\
   \bottomrule
\end{tabular}
\end{table*}

 \vspace{3pt}

\subsection{RQ2: Changing Patterns of Emotional Disclosure}
\label{section4.2}
We focused on how emotional disclosure changes during quarantine from long-term and period-based perspectives. More specifically, we first unveiled the changing patterns from April 2020 to July 2022 to have an overall exploration of emotions in this context, and then investigated those related to quarantine durations, which further understood the day-by-day emotions during quarantine.

\subsubsection{Long-term Changing Patterns}
\begin{figure}[htbp]
\centering 
\includegraphics[width=0.48\textwidth]{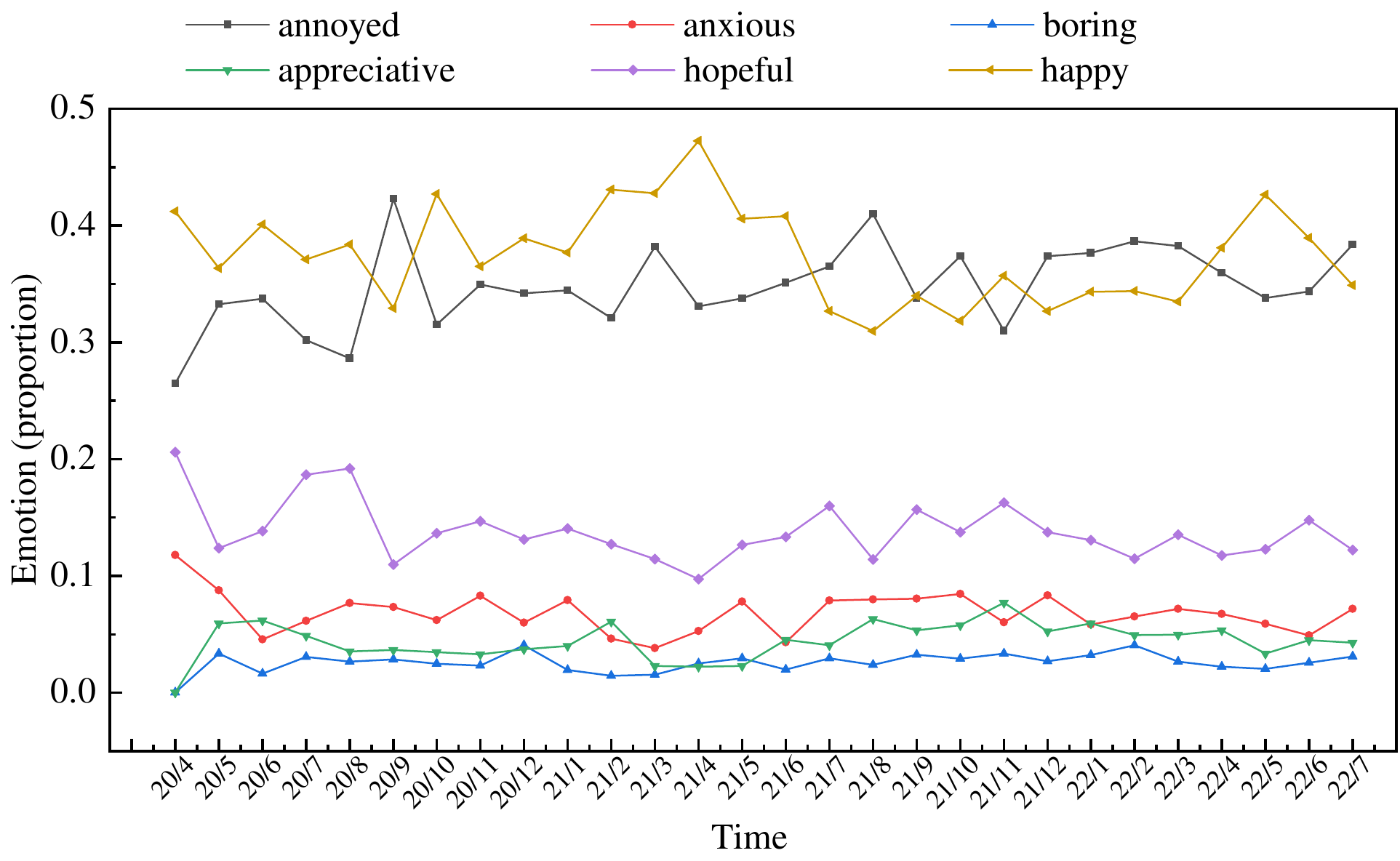} 
\caption{The long-term changing patterns of emotional disclosure from April 2020 to July 2022.} 
\label{figure3}
\end{figure}

Figure \ref{figure3} delineated the long-term changing patterns. Several general trends could be discovered: (1) The overall increase in the proportion of \emph{annoyed} indicated a growing trend of individuals expressing and emphasizing their annoyance over other emotions; (2) All emotion proportions gradually stabilized on the whole, probably due to the stability of the quarantine policy and the spread of the vaccine, which brought support to the public emotions; (3) \emph{Appreciative} and \emph{boring} proportions were very stable in the whole process.
\subsubsection{Period-based Changing Patterns}
Except for the temporal characteristics from a long-term perspective, changing patterns in different durations of quarantine were also noteworthy. In particular, we focused on the 21-day global changing patterns based on posts whose labels of quarantine day were within 21 days, prompting a comprehensive understanding of temporal characteristics as quarantine time went by. Then we uncovered 7-day and 14-day grouped changing patterns based on users whose quarantine durations were 7 days or 14 days, to unearth the similarities and differences of emotions in different quarantine durations.

\begin{figure}[htbp]
\centering 
\includegraphics[width=0.48\textwidth]{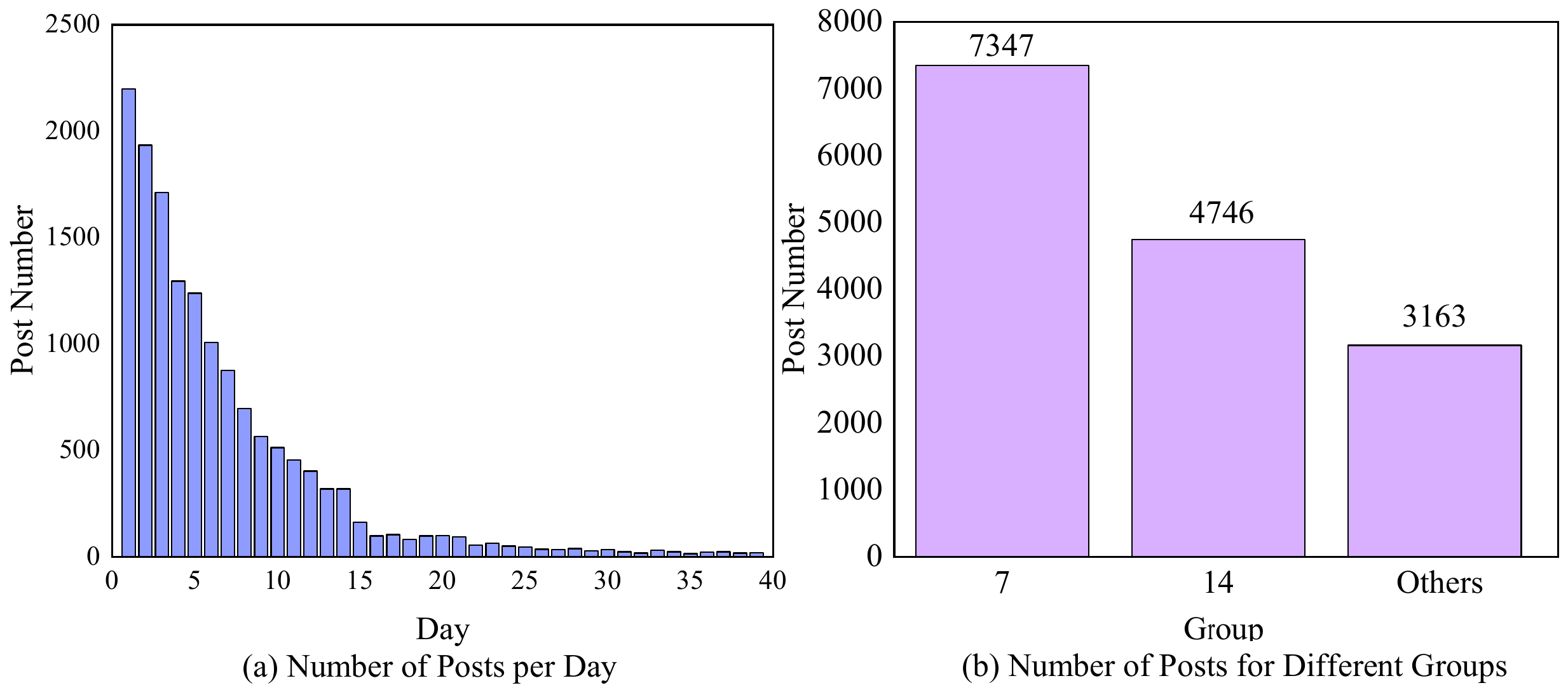} 
\caption{The statistical results of post number for (a) per day and (b) different groups (including 7-day quarantine group, 14-day quarantine group and the remaining group).} 
\label{figure4}
\end{figure}

\begin{figure*}[htbp]
\centering 
\includegraphics[width=0.94\textwidth]{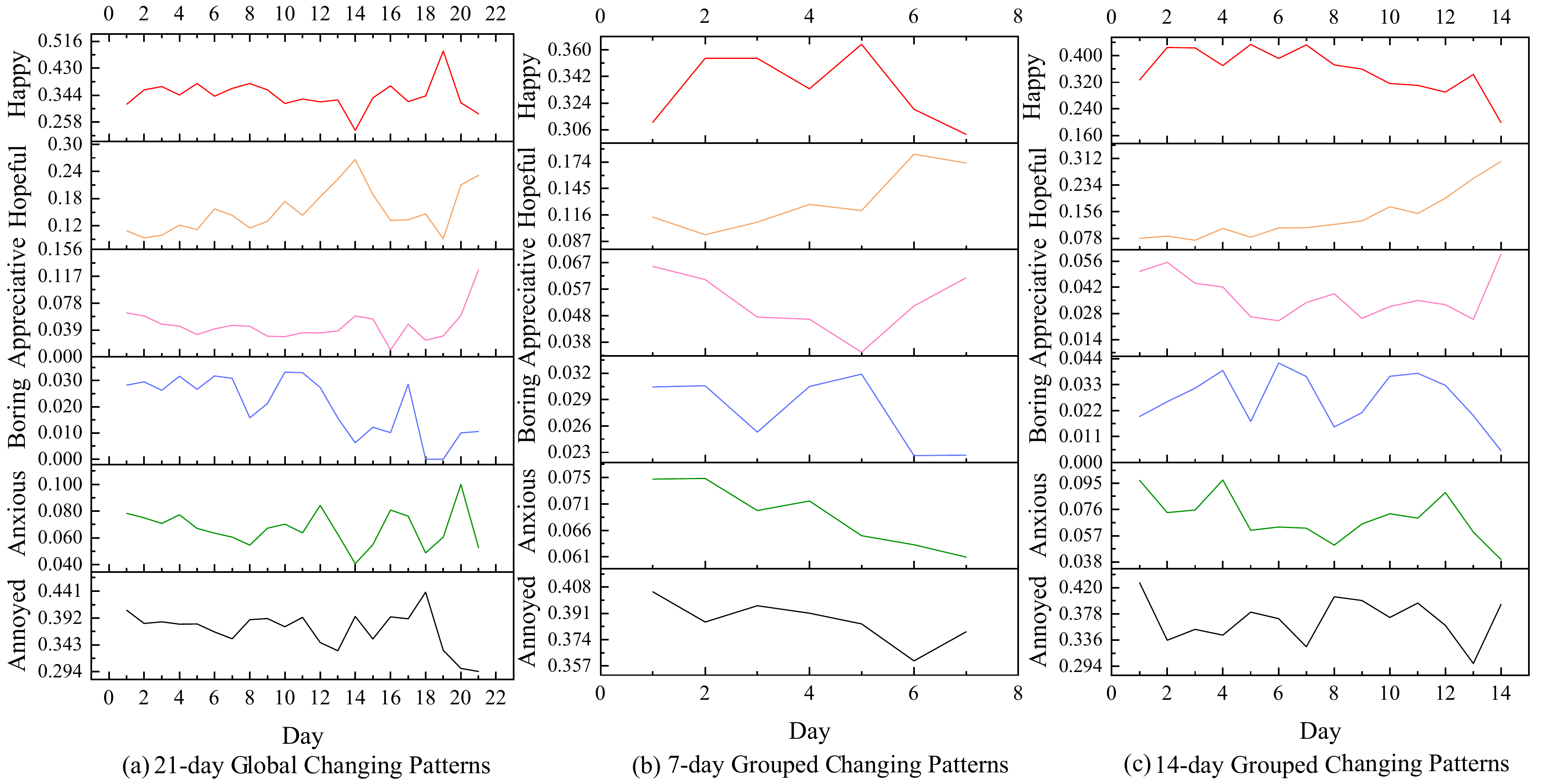} 
\caption{The period-based changing patterns from (a) 21-day global perspective, as well as (b) 7-day and (c) 14-day grouped perspectives.} 
\label{figure5}
\end{figure*}

\textbf{21-day global changing patterns:} To begin with, we statistically illustrated the trend of post number as the increase of quarantine duration. Based on Figure \ref{figure4} (a), we noticed that the post number after the 21st day was less than 100, not enough to implement a sufficient and adequate analysis, so we chose 21 days as the analysis range. Figure \ref{figure5} (a) demonstrated the changing patterns of emotional disclosure during a 21-day diary-keeping. The results showed people in quarantine were able to adjust their mental health to some extent. In particular, the negative emotional disclosure relatively occupied high positions at the beginning of quarantine, then gradually decreased, except for some fluctuation for \emph{anxious} later on. The gradual decrease in the proportion of negative emotions indicated that individuals in quarantine tended to achieve a psychological balance as they might gain a better understanding of the nature of the epidemic. Furthermore, positive emotions could also validate the idea: \emph{hopeful} as a whole were on the rise, especially near the end of quarantine (i.e., 14th, 20th, 21st days). The changing pattern of \emph{appreciative} showed that people maybe initially felt \emph{appreciative} of the support received from others, but as time went on, the novelty wore off and feeling of isolation might arise, contributing to a decrease. However, as the quarantine neared its end, appreciation could be rekindled since people built up anticipation and excitement, accompanied by renewed support and encouragement from their surroundings.


\textbf{7-day and 14-day grouped changing patterns:} Figure \ref{figure4} (a) delineated that the number of posts had continuously dropped for the first 14 days and stabilized thereafter, where the difference was relatively large for the first 7 days and around the 14th day. Figure \ref{figure4} (b) demonstrated that 7 days and 14 days were the dominant quarantine durations. The results were in line with the quarantine policy \cite{policy} which recommended 7 days and 14 days as quarantine or health-monitoring periods. The 7-day and 14-day diary-keeping of emotional disclosure were indicated in Figure \ref{figure5} (b) and (c). The results yielded some detailed and interesting characteristics: (1) On the 3rd and 4th day, \emph{anxious} and \emph{boring} were increasing and \emph{annoyed} maintained at a relatively high level, which suggested that these two days might have posed significant challenges for individuals; (2) \emph{Anxious} showed a pattern of first decline and then rise between the 4th day and 12th day, with a notable increase starting on the 8th day. It indicated that during shorter durations (e.g., from the 4th day to the 8th day), anxious emotions tended to be alleviated. However, for longer durations (e.g., from the 8th day to the 12th day), anxiety could become more prominent; (3) An interesting finding was that on the last day of the quarantine, \emph{annoyed} would upsurge from a relatively low proportion. This could be attributed to the common quarantine durations, where people might have anticipated the end of quarantine on either the 7th or 14th day, only to discover that the quarantine period was extended. The uncertainty and unexpected prolongation of the isolation period might lead to a higher level of annoyance among individuals.

The comparison between the two kinds of grouped changing patterns signified that although the overall trends of emotional disclosure were similar, there still existed differences during the same period to some extent. For instance, from the 4th to 6th day, 7-day quarantined people reported a decrease in \emph{annoyed}, possibly due to a sense of relief as their isolation period neared its end. However, individuals who were quarantined for 14 days experienced an increase in \emph{annoyed} on the 5th day. This increase could be attributed to the realization of a longer remaining isolation period, leading to frustration and irritation. Moreover, in terms of \emph{anxious}, individuals in 7-day diary-keeping stayed at a high level for the first 2 days while those in the 14-day grouped diary-keeping exhibited a decrease after the first day.

\subsection{RQ3: Latent Factors Contributing to Negative Emotional Disclosure}
\label{section4.3}
After keeping track of some interesting temporal characteristics of emotional disclosure, we probed further into the latent factors that might contribute to negative emotional disclosure, attempting to explore effective countermeasures. Table \ref{table2} and Table \ref{table3} summarized the topical themes and corresponding keywords with LDA after an inductive thematic analysis described in Section \ref{RQ3}. The abundance of topical themes indicated there were diverse possible factors denoting negative emotional disclosure, and the reasons behind each negative emotion were also extensive. To effectively and precisely alleviate negative emotions, managers of epidemic prevention and control were suspected to pay more attention to divergent and fundamental measures instead of purely providing psychological counseling to regulate emotions in quarantine.


\subsubsection{Latent annoyed Factors}
Based on the results of Table \ref{table2} and the corresponding posts, there were four topical themes related to \emph{annoyed}: (1) Individuals in quarantine were inclined to complain about the \emph{endless epidemic} about the frequent nucleic acid test requirements, a flurry of reports about the virus and confirmed cases, etc; (2) During quarantine, people felt physically and mentally \emph{restricted} as if they were in jail and longed for spirit disinhibition such as going out and enjoying the scenery of the world. Moreover, the \emph{solitude} property profoundly aroused a sense of loneliness, evoking the eagerness for companionship; (3) The \emph{management measures and sometimes disorderly arrangement} also irritated people in quarantine due to sporadic cases and the local implementation. It was argued that some local arrangements were suddenly changed and the attitude of services and efficiency of implementations could lead to annoyance; (4) Poor \emph{infrastructure} was also one of the major reasons for provoking exasperation, such as the accommodation, accessibility (e.g., takeout and express delivery), and supply or taste of food. 
\begin{table*}
\caption{List of topical themes and representative keywords with snippets of example posts in \emph{annoyed}.}
\label{table2}
\centering
\begin{tabular}{p{3cm}p{4cm}p{6cm}}
   \toprule
   Topical Theme & Keywords & Example\\
   \midrule
  endless epidemic & nucleic acid (test), epidemic situation, notification, lifting (lockdowns), diagnosis, report, global, virus, release, information & \emph{I have to do the nucleic acid test every day, and can’t even attend the class. The epidemic is really annoying. When will this epidemic end?}\\
  \cline{2-3}
  freedom restriction and solitude & expect, monotony, go out, accompany, world, scenery, release from prison, closure, spirit, liberation & \emph{Tomorrow is another week, but I still don't know when I can go out.}\\
  \cline{2-3}
  management measures and disorderly arrangement & policy, rule, management, street, service, measure, adjustment, wait, postpone, efficiency & \emph{. . . Why is the policy so rigid. . . Not everyone can sit at home all the time without earning money. . . How can there be different quarantine methods in the same area? When consulting the policy, no department can give a reasonable explanation or document.}\\
  \cline{2-3}
  infrastructure (food, accommodation) & room, boxed meal, takeaway, quilt, bread, window, express delivery, disinfect, vegetables, eggs & \emph{As soon as I entered the room, I was stunned. A lot of hair on the ground and unknown marks on the pillow and bed. Dirty and messy!}\\
   \bottomrule
\end{tabular}
\end{table*}

\subsubsection{Latent anxious Factors}
Table \ref{table3} signified the following five latent \emph{anxious} factors: (1) Quarantine had \emph{disrupted our lives}, both in terms of physical adaptation (e.g., internal clock) and mundane goings-on of their lives (e.g., work and study); (2) Despite being in quarantine, they still \emph{worried about being infected} during 
transit from home to quarantine locations, on the way to the nucleic acid test, or due to the confirmed cases around (e.g., communities and companies); (3) A sense of apprehensiveness might be engendered on account of the \emph{uncertainty of influences by management measures}. For example, the public did not know when and how long they would be in quarantine. An unexpected finding was that a knock on the door made people feel anxious; (4) Individuals felt anxious about both the \emph{mental and physical health}. The state of mind was easy to break down and insomnia was a normal phenomenon during quarantine. In addition, the illness rather than COVID-19 (e.g., common cold or fever) was very anxiety-provoking; (5) The \emph{supply} of food was of great importance, in particular for people in home quarantine who had to vie for food on online food delivery apps usually with a high price, also causing concern of finance. 
\begin{table*}[htbp]
\caption{List of topical themes and representative keywords with snippets of example posts in \emph{anxious}.}
\label{table3}
\centering
\begin{tabular}{p{3.1cm}p{3.9cm}p{6cm}}
   \toprule
   Topical Theme & Keywords & Example\\
   \midrule
  disturbance of daily routine & wake up, internal clock, adjustment, job, work, online course, work from home, study, punch in, homework & \emph{My internal clock is completely messed up. I don’t want to sleep every night and can’t wake up in the morning. The worst thing about quarantine is the exhaustion of mental energy, and the living space, workspace and study space are very chaotic.}\\
  \cline{2-3}
  uncertainty of infection & virus, protective clothing, contact, nucleic acid, infection, diagnosis, transit, unit, community, positive & \emph{It is hoped that the process of nucleic acid test can be more standardized so that there will be no more cross-infection.}\\
  \cline{2-3}
  uncertainty of influences by management measures & staff, concentrated (quarantine), knocking, government, accommodation, inn, ambulance, hotel, home, reception & \emph{I used to feel the epidemic was far away from me, but this time I really feel the fear. One is the anxiety of waiting and not knowing when I will be taken away for quarantine. The other is...}\\
  \cline{2-3}
  health & insomnia, see a doctor, sleepless, cough, fever, throat, sickness, state of mind, spirit, breakdown & \emph{My head is about to explode probably because the hotel is so cold and I didn't dry my hair after washing.}\\
  \cline{2-3}
  supply & starve to death, steamed buns, food, rice, instant noodles, grocery shopping, refrigerator, sustenance, eggs, stock & \emph{Insomnia until six o'clock in the morning...who knows...I went to [a food delivery app] to grab food at six o'clock and didn't get it.}\\
   \bottomrule
\end{tabular}
\end{table*}
Based on these potential factors of negative emotional disclosure, a series of strategies and countermeasures could be considered by governments, Centers for Disease Control and Prevention (CDC) and other official organizations. For instance, governments could provide basic food, common medicines, psychological counseling and free online consultation to deal with \emph{unsatisfied infrastructure} and \emph{health concerns}; the staff could try to ensure that they knock on the door at a fixed time period to relieve the anxiety caused by \emph{uncertainty of management measures}; during the nucleic acid test, dividing people in quarantine into batches or taking other preventive measures to avoid cross-infection could be considered, in order to reduce \emph{uncertainty of infection}; schools or companies would adopt specialized management requirements for personnel in quarantine, reducing the stress of \emph{disturbance of daily routine}.

\section{DISCUSSION}
In this section, we reflect on our findings and enlighten practical and design implications to drive the understanding of emotional disclosure on social media in quarantine. First, we probe into the responses of people in quarantine under the emotional challenges, and propose implications for better balancing emotions in Section \ref{section5.1}. Then, we highlight design implications on social media and reflect on the health interventions in Section \ref{section5.2}. Afterwards, we think over how latent factors of negative emotions enlighten social support to mitigate negative psychological impacts in Section \ref{section5.3}.


\subsection{Immersed in Emotional Situation: Responses and Implications}
\label{section5.1}
By identifying quarantine-specific emotional categories, this work enriches the understanding of emotional disclosure from the perspectives of how individuals respond to quarantine, and how the responses demystify the practical and design implications to aid psychological balance during quarantine.

First, this work revealed the resilience and adaptivity of the public in the scenario of quarantine to some degree. As shown in Section \ref{section4.1}, the proportion of positive emotions was slightly higher than negative ones, and \emph{happy} (35.96\%) played the most significant role in positive emotions, indicating the emotional well-being of the population on average was challenging but not as bad as imagined. Meanwhile, our findings in Section \ref{section4.2} further demonstrated that negative emotions diminished over time. Indeed, a type of natural recovery capability (i.e., resilience and adaptivity) was corroborated. It also dovetailed with the cognitive dissonance theory \cite{festinger1962theory}, pointing out that in order to reduce negative psychological effects, people are more likely to change their cognition if there is no way of changing the environment. In this regard, adding the recovery capability as a reference index during the psychological treatment process in quarantine, would be promising to better help mental health management. 

Second, this work also uncovered severe psychological stress in people during quarantine. As demonstrated in Section \ref{section4.1}, the proportion of posts with negative emotions was 45.77\% among all posts. The massive negative content on social media could form a vicious circle like pyramid schemes. From the perspective of receivers, it is common that people have more engagement in negative content on social media \cite{peng2019discovering,naveed2011bad}. Negativity bias theory \cite{adikari2021emotions} indicates that people's behavior and cognition are more likely to be affected by negative emotions. From the other perspective of posters, on account of social compensation theory \cite{valkenburg2007preadolescents}, people with psychological distress have more intense motives to disclose on social media who desire more social connection and affiliation \cite{lee2013lonely, tian2013social,hollenbaugh2014facebook}. Furthermore, recall that the most predominant negative category was \emph{annoyed} exhibited in Section \ref{section4.1}, rather than \emph{anxious}. The more outstanding performance of \emph{annoyed} compared with \emph{anxious} bears out the fatigue phenomenon during quarantine due to endless epidemic \cite{williams2021variant,ross2021household}. Angry speech is proven to be more likely to receive abusive language, in turn, \cite{mechkova2021norms}. Then, a new round of negative outcomes emerges on social media. Thus, detecting and moderating negative emotional disclosure on social media is of great imperative, especially with the extensive negative exposure to social media during quarantine. In light of the fatigue phenomenon, how to adjust social support measures accordingly (e.g., maybe from previously disseminating professional information about the virus and debunking misinformation to providing more emotional accompany), deserves more in-depth exploration. Moreover, the emotion analysis approach in our work, which focuses on identifying quarantine-specific emotional categories, investigating emotional changing patterns and exploring latent factors of emotions, could be considered as an adjunct in mental health diagnosis during quarantine or subsequent epidemic era. Taking a broader perspective, our work offers compelling evidence supporting the feasibility and significance of conducting emotion analysis in affective computing by annotating context-specific emotions. By focusing on the emotions experienced within particular contexts, we can gain valuable insights into individuals' mental well-being and tailor interventions and support accordingly.
\subsection{Driven from Diary-keeping: Designs and Interventions}
\label{section5.2}
In the specific setting of quarantine, we notice a novel form of emotional disclosure: people in quarantine spontaneously write diaries on social media for mundane recording, experience sharing and opinion expression. By uncovering the diary-keeping behaviors on social media, we reflect on the design implications of how diaries during extraordinary times could be supported via social media, and the practical implications of health intervention strategies based on emotional disclosure.


The intimacy with data is well exhibited via social-media-based diary-keeping in our work. The narration technique on social media, leveraged by users to delineate their daily life, provides a complementary perspective for application-centered numerical tracking techniques (e.g. Personal Informatics apps \cite{cho2022reflection}) in the way of person-centered instructive description. The functions on social media like sharing, forwarding, commenting and mentioning (i.e., @) provide a platform for users to communicate with each other, stimulating and boosting the social and cultural potential. The changing patterns in Section \ref{section4.2} provided evidence that employing large-scale observation of quarantine diaries could, to some degree, serve as a viable approach for identifying and comprehending emotional disclosure. On this note, emotion-detecting systems and system-level interventions (e.g., reverse recommendation of positive information when detecting extreme negative emotions) could come into being by future designers and researchers. Note that we do not instigate users to open their private and specific data on social media but just encourage the narrative way of diaries. On the long view, the present identification approach of diary-self-tracking is based on the keyword (e.g., \emph{Quarantine Diary}) and temporal label (e.g., Day 1) which is manually tracked by users. To this end, system-facilitated construction of diaries (e.g., providing an elaborately-designed diary template) may be considered for future researchers. Also, integrating diary apps or other self-tracking functions into social media, and whether the diary interface should become more strict like numerical recording or more flexible like paper journaling, are a worth-exploring research direction. Nevertheless, some studies have shown that self-disclosure on social media, especially negative sharing, can engender a lot of risks (e.g., privacy) \cite{bazarova2014self} and need to be carefully created and negotiated \cite{andalibi2018responding}. Social media posts are also subject to content moderation and surveillance, so it is hard to assume that people share experiences and opinions of quarantine without concern.

Drawing on the outcomes of social-media-based diary-keeping, health intervention strategies are engendered. As shown in Section \ref{section4.2}, the discrepancy of each group in emotional disclosure demonstrated individuals might have different psychological statuses and concerns at the same period. Such different patterns enable governments to comprehend the people in quarantine from multiple perspectives when making healthcare policies, and specialized healthcare measures for each group are recommended. Meanwhile, health intervention strategies (e.g., psychological crisis consultation time) are suspected to dynamically adjust and focus on several days, such as the 3rd and 4th days when all negative emotions were at a high level. A bounce came up in \emph{annoyed} near the end of quarantine, indicating the health management department cannot ignore the negative emotions close to the end. A plausible interpretation is that approaching the end of quarantine, people are suddenly informed to remain in quarantine with the \emph{annoyed} increasing. These findings empirically validated the feasibility of understanding emotional disclosure from a quantitative perspective. The changing patterns can also enlighten mental health regulations during similar public health emergencies.



\subsection{Enlightened by Latent Factors of Negative Emotional Disclosure: Insights on the Social Support}
\label{section5.3}
Prior work ascribed great importance to self-disclosure and social support as critical approaches to improving living standards and self-efficacy. Negative emotional disclosure demonstrates the latent concerns and dissatisfaction of the public and functions as an impetus to provide available direction for social support. This section further discusses what kind of social support could be provided to alleviate negative emotions based on self-disclosure.

This work comprehensively signified the intricate characteristics of factors that might account for negative emotions and revealed the potential need for social support. Driven by the social support theory \cite{cohen1985stress}, and the effect mechanism of self-disclosure and psychological well-being \cite{luo2020self}, typical social support categories consist of \emph{instrumental} (i.e., the enacted supportive behaviors) and \emph{emotional} (i.e., the perceived impression) support. In essence, our observations of potential social support needs in Section \ref{section4.3} were in line with the theory-driven category. On the one hand, \emph{instrumental} social support to deal with practical difficulties like \emph{short supply} and \emph{poor infrastructure} from governments and related departments, is imperative. Also, given that \emph{uncertainty} is a remarkable factor during quarantine, construing a channel on social media to interpret policies and measures, may be beneficial to mitigate negative emotions and avoid more serious situations caused by misinformation. On the other hand, social media together with official organizations, could be taken into account for \emph{emotional} social support. The emotional needs like \emph{freedom restriction and solitude} and \emph{feeling the epidemic is endless}, verified online accompany is vital due to abrupt reduction of physical interactions. On this note, tailored accompanying communities consisting of experienced people, professionals or digital volunteers, and interactive chatbots on social media are recommended for future designers to cure people in quarantine who are grappling with negative emotions.

\subsection{Limitations and Future Work}
This work adopted a quantitative approach to comprehensively investigate the categories of emotional disclosure, the changing patterns of emotional disclosure and factors that might give rise to negative emotions during quarantine, which enlighten the practical and design implications for better relieving mental health issues. Nonetheless, this work has the following limitations: (1) While social media can serve as a means of communication, posts shared on social media may not completely capture or convey the full range of emotions or explicit and implicit needs of the public. Individuals moderating their content and engaging in self-surveillance on social media could present a curated version that doesn't reflect true emotions or comprehensive needs. (2) Due to the presence of multiple or repeated types of temporal information within posts, the exact accuracy of the extracted temporal information cannot be guaranteed, even though we have employed rules based on manual encoding and extensive experience.

When keeping a diary on social media becomes prevalent but related interface designs are still limited, we encourage a customized diary interface while combining the features of personalization and flexibility on social media in the future. Additionally, future work could focus on the difference between social-media-based diaries and paper journaling to obtain valuable inspiration for a user-friendly interface.

\section{CONCLUSION}
In this paper, we explore the categories of emotional disclosure during quarantine, the changing patterns of emotional disclosure and possible factors that might contribute to negative emotions through a quantitative approach. By investigating the posts during quarantine on a popular Chinese social media platform Weibo, we identify emotional categories in quarantine, and discover the distribution of different emotions. The results demonstrate that posts with negative emotions are non-negligible which approximately account for half of all posts. Furthermore, the changing patterns of emotional disclosure indicate the resilience of the public with a declining trend of negative emotions during quarantine, yet the annoyed emotions upsurge near the end of the quarantine period. Finally, we uncover the possible influencing factors of negative emotions like disturbance of daily routine, poor infrastructure as well as the uncertainty of influences by management measures. Based on the findings, we propose practical and design implications to ease the psychological stress during quarantine.
\begin{acks}
The research was supported in part by a RGC RIF grant under the contract R6021-20, RGC CRF grants under the contracts C7004-22G and C1029-22G, and RGC GRF grants under the contracts 16209120, 16200221 and 16207922.
\end{acks}



\bibliographystyle{ACM-Reference-Format}
\bibliography{references}
\end{document}